\documentclass[manuscript]{aastex6}

\newcommand{\Alfven}{Alfv\'{e}n\ }

\newcommand{\sdo}{{\it SDO}}

\newcommand*{\figw}[1]{Figure~\ref{#1}} 
\newcommand*{\figs}[1]{Figures~\ref{#1}}

\newcommand*{\sect}[1]{Section~\ref{#1}}

\newcommand*{\unit}[1]{\ensuremath{\mathrm{\, #1}}}
\newcommand*{\cm}{\unit{cm}}
\newcommand*{\km}{\unit{km}}
\newcommand*{\ps}{\unit{s}^{-1}}   
\newcommand*{\kmps}{\km \ps}
\newcommand*{\pcms}{\unit{cm}^{-2}}	
\newcommand*{\pcmc}{\unit{cm}^{-3}}	

\newcommand*{\ergs}{\unit{ergs}}
\newcommand*{\s}{\unit{s}}

\newcommand*{\E}[1]{\times 10^{#1}}

\shorttitle{Counter propagating fast magnetosonic wave trains}
\shortauthors{Ofman and Liu}

\begin{document}

\title{Quasi-periodic Counter-propagating Fast Magnetosonic Wave Trains from Neighboring Flares: 
 SDO/AIA Observations and 3D MHD Modeling}
\author{Leon Ofman\altaffilmark{1,2,3}, Wei Liu\altaffilmark{4,5,6}}	
\altaffiltext{1}{Catholic University of America, Washington, DC
20064} \altaffiltext{2}{NASA Goddard Space Flight Center, Code 671,
Greenbelt, MD 20771} \altaffiltext{3}{Visiting, Department of
Geosciences, Tel Aviv University, Tel Aviv, Israel}
\altaffiltext{4}{Bay Area Environmental Research Institute, NASA Research Park, Building 18, Mailstop 18-4, Moffett Field, CA 94035-0001, USA}
\altaffiltext{5}{Lockheed Martin Solar and Astrophysics Laboratory, 3251 Hanover Street, Bldg. 252, Palo Alto, CA 94304, USA}
\altaffiltext{6}{Visiting Scholar, W. W. Hansen Experimental Physics Laboratory, Stanford University, Stanford, CA 94305, USA}

\begin{abstract}
Since their discovery by \sdo/AIA in EUV, rapid (phase speeds of $\sim$1000 km s$^{-1}$), quasi-periodic, fast-mode propagating wave trains (QFPs) have been observed accompanying many solar flares. They typically propagate in funnel-like structures associated with the expanding magnetic field topology of the active regions (ARs). The waves provide information on the associated flare pulsations  and the magnetic structure through coronal seismology. The reported waves usually originate from a single localized  source associated with the flare. Here, we report the first detection of counter-propagating QFPs associated with two neighboring flares on 2013 May 22, apparently connected by large-scale, trans-equatorial coronal loops. We present the first results of 3D MHD model of counter-propagating QFPs an idealized bi-polar AR. We investigate the excitation, propagation, nonlinearity, and interaction of the counter-propagating waves for a range of key model parameters, such as the properties of the sources and the background magnetic structure. In addition to QFPs, we also find evidence of trapped fast (kink) and slow mode waves associated with the event. We apply coronal seismology to determine the magnetic field strength in an oscillating loop during the event. Our model results are in qualitative agreement with the AIA-observed counter propagating waves and are used to identify the various MHD wave modes associated with the observed event providing insights into their linear and nonlinear interactions. Our observations provide the first direct evidence of counter-propagating fast magnetosonic waves that can potentially lead to turbulent cascade and carry significant energy flux for coronal heating in low-corona magnetic structures.
\keywords{Sun:~activity --- Sun:~corona --- Sun:~oscillations --- Sun:~flares --- waves --- magnetohydrodynamics (MHD)}
\end{abstract}

\section{Introduction}
The Atmospheric Imaging Assembly (AIA) \citep{Lem12} on board the {\it Solar Dynamics Observatory} ({\it SDO}) has provided unprecedented high-resolution, high-cadence, nearly continuous view of the full-Sun corona in Extreme Ultraviolet (EUV) since its launch in 2010. This allows routine detection of many types of coronal EUV waves  \citep[see, e.g., the review by][]{LO14}. In particular, since their initial detection \citep{Liu10,Liu11} and identification with 3D MHD modeling \citep{Ofm11}, quasi-periodic fast propagating (QFP) magnetosonic wave trains are now routinely observed to accompany flares \citep[e.g.,][]{Liu12,She12,She13,she17, She18,Yua13, Kum13, Nis14,Zha15,God16,Qu17}. The QFP waves can carry significant energy fluxes (on the order of $10^5$~erg~cm$^{-2}$~s$^{-1}$) needed to heat the active-region (AR) corona. However, the contribution of these events to coronal heating is probably small due to their infrequent occurrences, based on present observations. Nevertheless, many such QFP wave trains have been observed and are now available in a statistical study of the order of 100 events \citep{Liu16}. Numerical modeling of QFPs was first performed using 3D magnetohydrodynamics (MHD) in a bipolar AR model \citep{Ofm11}, and was later studied using 2D MHD models as waves trapped in an MHD wave-guide \citep{Pas13,Pas17} or generated by quasi-periodic perturbations resulting from magnetic reconnection \citep[e.g.,][]{Yan15,Tak16,Tak17}.

The propagation and phase speed of QFP waves provide information on the magnetic structure of ARs, and it has been shown that the quasi-periodicity is directly related to flare pulsations \citep[e.g.,][]{Liu11} and to radio bursts \citep[e.g.,][]{Yua13,God16,Kum17}. The importance of QFPs in coronal diagnostics is rooted in the direct connection that they provide between the driving solar flare or coronal mass ejection (CME) and the underlying AR magnetic structure as inferred from their co-temporal evolution and co-spatial origin. In particular, QFPs can shed light on the long-standing puzzle of flare pulsations seen in a wide range of wavelengths from radio to hard X-rays. Moreover, these fast mode waves are propagating, thus, sampling the coronal magnetic field over a significant spatial extent in an AR and beyond. The events that produce QFPs often induce damped coronal loop oscillations of large amplitudes that can be used for coronal seismology (CS) \citep[e.g., see the reviews][]{NV05,LO14,Wan16} of the closed loop structures in the AR magnetic field. 

In the present paper we report the first observation of counter-propagating QFP wave trains that appear to interact and are associated with coronal loop oscillations. There are only few observations of counter-propagating fast mode waves in the low corona with potential evidence  in transequatorial loops \citep[e.g.,][]{Tom09,Dem14}.  Related counter-propagating Alfv\'{e}n waves, which could be associated with fast magnetosonic waves in inhomogeneous structures, are often invoked for the nonlinear interaction and generation of turbulence in incompressible  plasma that leads to energy cascade towards small scales and eventually to plasma heating. Observational evidence for such counter-propagating waves has been found in the solar wind \citep[e.g.,][]{Tu89,Bru97,BC05}.

The observational analysis of counter-propagating QFPs and associated wave modes provide motivation for the numerical 3D MHD modeling of counter-propagating QFP wave trains in an idealized, but general bipolar AR model. The goal of the model is to identify and investigate the various wave modes in the observation with coronal seismology applications (coronal wave heating is not modeled explicitly here). The model provides insights on the MHD mode couplings due to linear and nonlinear effects, for several scenarios of magnetic geometry and wave sources. The paper is organized as follows: in Section~\ref{obs:sec} we present the \sdo/AIA observations, analysis of the counter-propagating QFP wave trains, and transverse loop oscillations; in Section~\ref{mhd:sec} we describe the details of the MHD model of the counter-propagating QFP wave trains; Section~\ref{num:sec} is devoted to numerical results of the model; and finally the conclusions and discussions are presented in Section~\ref{disc:sec}.

\section{\sdo/AIA Observations and Analysis}
\label{obs:sec}

On 2013 May 22, an M5.0 class flare occurred in NOAA AR~11745 (N13W75) at 12:35~UT and peaked at 13:32~UT in {\it GOES} soft X-ray flux (see \figw{spacetime:fig}(a)). This flare was associated with a major eruption involving a flux rope \citep{LiT13,Che14} and a fast CME at a speed of $\sim$$1400 \,\,\rm{km~s} ^{-1}$ \citep{Din14}. This eruption also produced a global coronal EUV wave. About 25~minutes after the onset of this flare, another, secondary flare occurred at 13:00~UT in the neighboring AR~11746 (S20W75) across the equator. This secondary flare could potentially be a sympathetic event following the first flare, because of their close timing and spatial separation \citep[e.g.,][]{PH90}. Both flares produced QFP waves, some of which appear to be propagating in opposite directions along the large-scale, trans-equatorial loops that connect the two flares. The interaction between these counter-propagating fast-mode wave trains and the properties of the propagating and trapped waves are the main focus of this study. 



\begin{figure*}[ht]
\center
\includegraphics[angle=90,width=14cm]{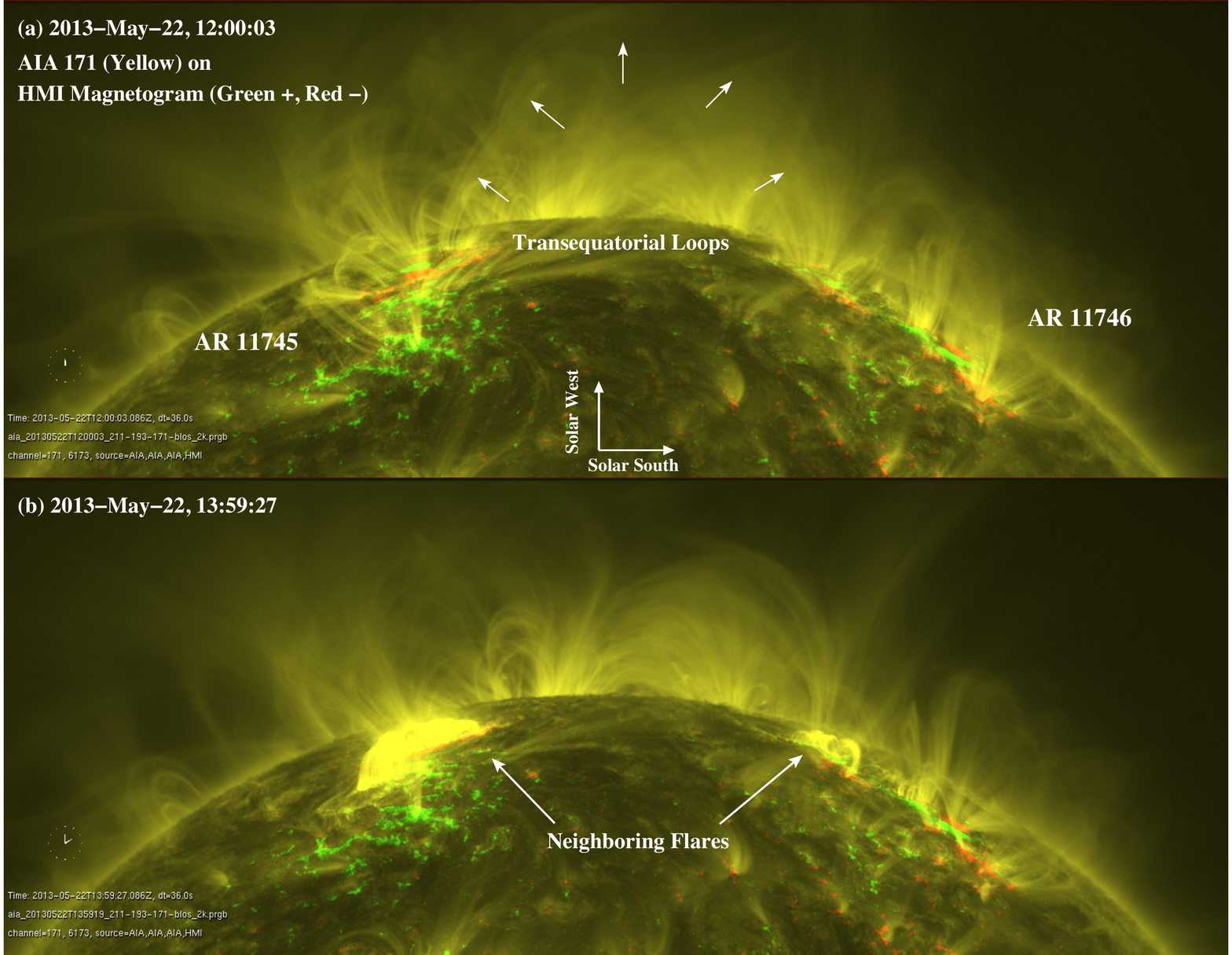}
\vspace{-2cm}\includegraphics[angle=90,width=14cm]{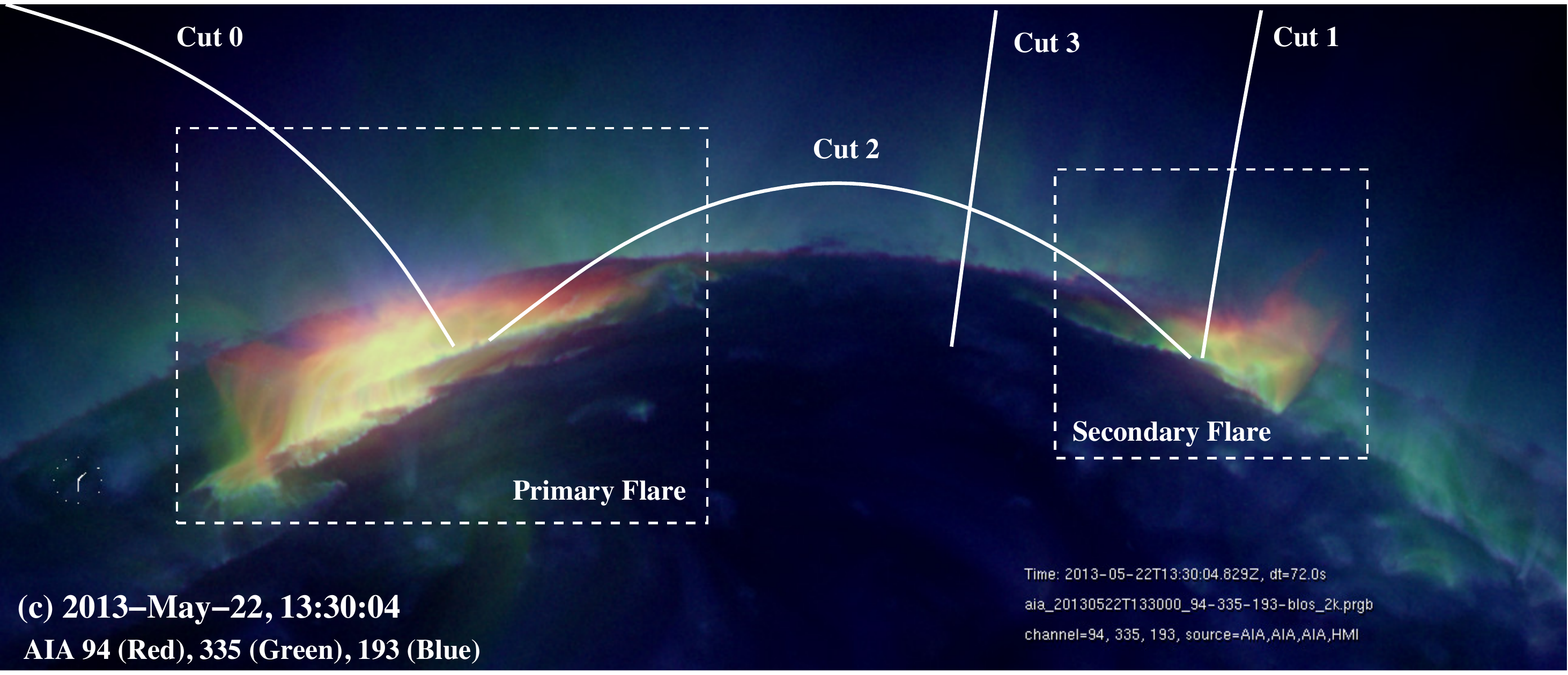}
\caption{Context \sdo/AIA images of the 2013 May 22 M5.0 flare/QFP event, rotated by 90 degrees counter-clockwise with the solar west up.
(a) Pre-flare 171~\AA\ image (yellow) at 12:00~UT overlaid on concurrent HMI magnetogram (green and red for positive and negative polarities, respectively). The short arrows mark the trans-equatorial loops that connect the two ARs, 11745 and 11746. (b) Same as (a) but later at 13:59~UT showing the neighboring flares, where QFP wave trains originate. (c) Composite image of three AIA channels of decreasing characteristic temperatures: 94~\AA\ (red), 335~\AA\ (green), and 193~\AA\ (blue) at 13:30~UT, when the flare is most evident in the hottest channel (red). We mark four spatial cuts, 0\,--\,3, used to obtain space\,--\,time plots and two boxed regions used to obtain the individual flare light curves, both shown in \figw{spacetime:fig}. (Animations 1 and 2 corresponding to the top panel are available in the online article.)
}
\label{2flares:fig}
\end{figure*}

Figure~\ref{2flares:fig} shows context images of this event from \sdo/AIA. The trans-equatorial loops that connect the two flaring sites between opposite magnetic polarities%
\footnote{Because of projection effects at these near-limb locations, the line-of-sight (LOS) magnetograms shown here may not represent the true magnetic polarities. However, these trans-equatorial loops were also evident a few days earlier when the two ARs were on the disk far from the limb with minimum LOS effects (see, e.g., \href{http://sdowww.lmsal.com/suntoday/index.html}{http://sdowww.lmsal.com/suntoday/index.html}).}
can be seen in panel~(a) at 12:00~UT, just before the flare onset. This is the observational basis for our adoption of a dipole geometry in our simulation described in \sect{mhd:sec}. In \figs{2flares:fig}(b) and (c), the two flares at later times (13:59 and 13:30~UT) during the gradual phase are evident, with the arcade of bright loops fully developed and hot emission above it best seen in the AIA 94~\AA\ channel in red (panel (c)). Note that the trans-equatorial loops are not obvious in panel~(b), which is likely a result of EUV dimming due to the CME eruption \citep[e.g.,][]{Zar99}. The evolution of this event can be best seen in the animations available in the online article.

\begin{figure*}[ht]
\center
\includegraphics[angle=90,width=\textwidth]{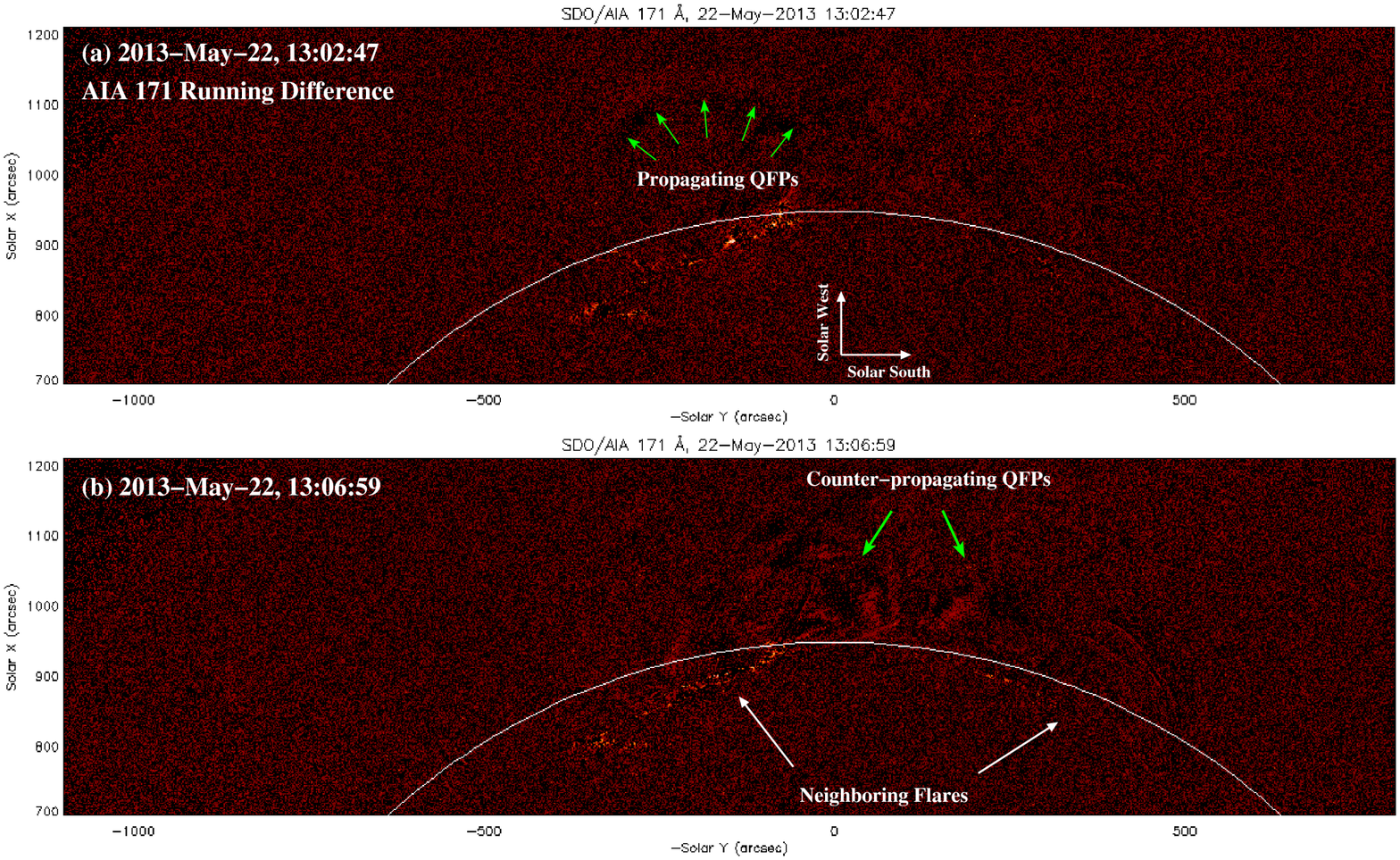}
\caption{
Running difference \sdo/AIA 171~\AA\ images on 2013 May 22 rotated with the solar west up showing (a) outward propagating QFPs from the primary flare, and (b) counter-propagating QFPs (apparently along trans-equatorial loops) from the primary flare on the left and the secondary flare on the right. (An animation of this figure is available online.)
}
\label{diff2flares:fig}
\end{figure*}

The QFP waves are most evident in AIA 171~\AA\ running difference images, as shown in Figure~\ref{diff2flares:fig}. Panel~(a) shows an example of QFPs propagating outward from the primary flare in AR~11745 early on at 13:02~UT. After the second flare develops, QFP waves propagating in opposite directions apparently along the marked trans-equatorial loops between the two flares become more evident. Panel~(b) shows an example of two counter-propagating wave-fronts at 13:06~UT just before they meet and apparently interact. Partly because of their high speeds and coherence over a large distance range, the QFPs are best seen in the online animation accompanying \figw{diff2flares:fig}. They appear to have similar wavelengths and amplitudes between the counter-propagating wave fronts. In addition to QFPs, other changes in the surrounding corona, as well as signatures  of trapped waves in the adjacent loops are evident.
 In particular, transverse oscillations of active region loop bundles can be seen in the animations accompanying Figure~\ref{2flares:fig}, marked with an arrow there (see below for their detailed analysis).  We also find EUV brightenings traveling along the loop at times, which we interpret as likely evidence of slow-mode waves coupled with kink oscillations that cause further density, temperature, and thus EUV intensity variations. This interpretation is supported by the 3D MHD modeling results, presented in \sect{mhd:sec} below.


While it is possible that the two QFP waves are independent of each other and their apparent interaction is a result of LOS integration of optically-thin EUV emission, it is very likely that the interactions are real as supported by (i) the presence of trans-equatorial loops and evidence of magnetic connectivity, (ii) the exactly opposite directions of propagation of these wave fronts with narrow angular extents, (iii) the space\,--\,time analysis results (\sect{ST:sec}), and (iv) the 3D MHD modeling (\sect{mhd:sec}) results presented below. 
%

\subsection{Space-time analysis}
\label{ST:sec}

We performed space\,--\,time analysis of the 2013 May 22 double flare/QFP event using various spatial cuts on \sdo/AIA images (indicated on Figure~\ref{2flares:fig}(c)) and determined the propagation speeds of the QFPs, as well as the oscillation properties of the adjacent loops.

\begin{figure}[ht]
\center
\includegraphics[width=13cm]{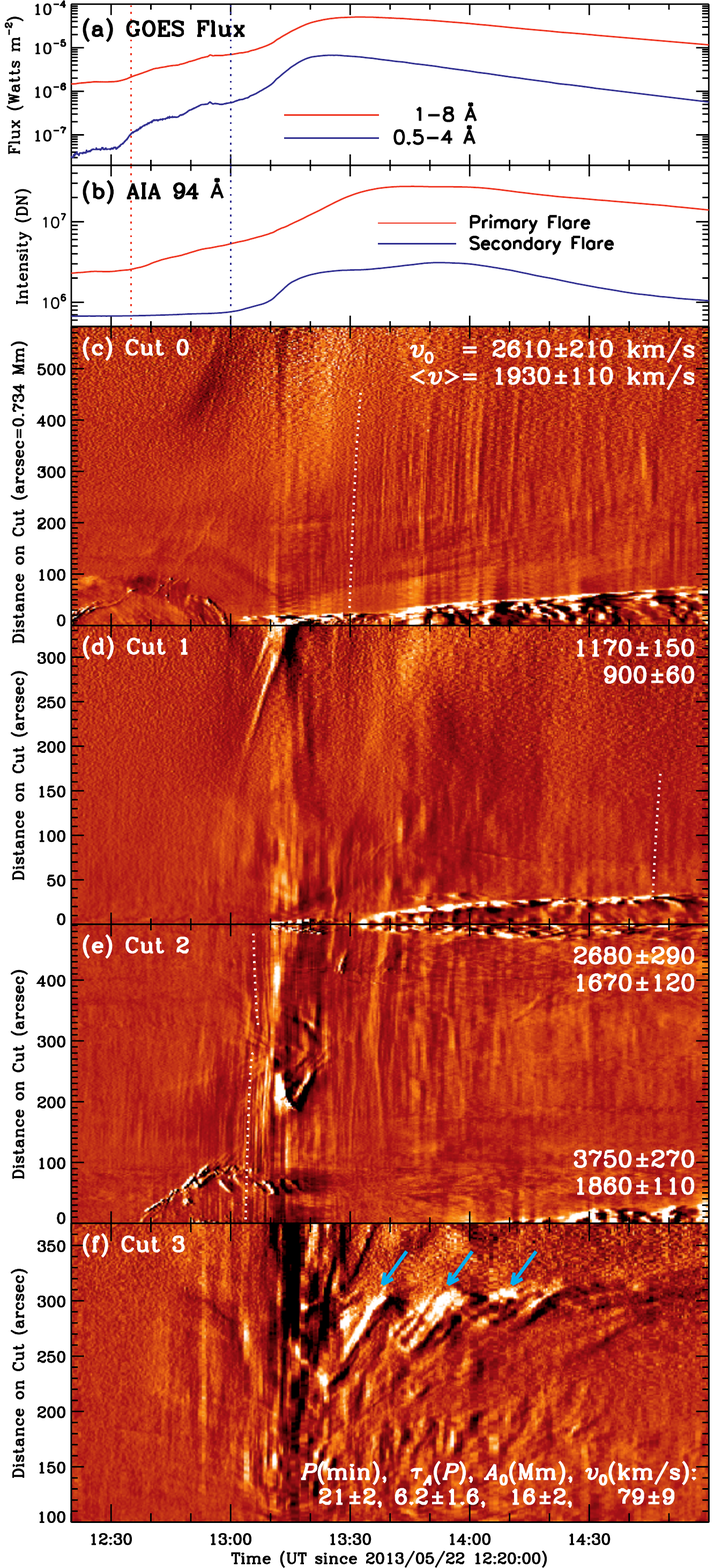}
\caption{\scriptsize	
Temporary evolution of the 2013 May 22 flares and QFP waves. (a) {\it GOES} soft X-ray fluxes. (b) AIA 94~\AA\ intensities of the primary (red) and secondary (blue) flares integrated over the boxed regions shown in Figure~\ref{2flares:fig}(c). The onsets of the two flares are marked by vertical dotted lines of the corresponding colors. The next four panels show running ratio space--time plots of the AIA 171~\AA\ channel from the four cuts shown in Figure~\ref{2flares:fig}(c). QFPs appear as steep narrow stripes, with primarily positive slopes for outward propagating waves shown in (c) and (d) from a region that appears to correspond to open magnetic funnel Cuts~0 and 1 rooted at the two flares, and both positive and negative slopes for counter-propagating waves shown in (e) from loop Cut~2 connecting the two flares. Distance in panel (e) is measured from the northern, main flare site. White dotted lines in (c)--(e) are parabolic fits to selected wave pulses, labeled with the initial speed $v_0$ from a parabolic fit and the average speed $\langle v \rangle$ from a linear fit in km s$^{-1}$.  
 (f) Space--time plot from the vertical Cut~3 showing damped transverse oscillations of the trans-equatorial loops. 
The best-fit parameters of a typical oscillation are labeled at the bottom (from Figure~\ref{model_vs_obs:fig}(c)). The periodic  loop intensity variations marked with the cyan arrows may contain contribution from the slow mode waves as suggested by the modeling results below.
}\label{spacetime:fig}
\end{figure}

Figures~\ref{spacetime:fig}(a) and (b) show flare light curves to provide a timeline context. Panel~(a) shows the {\it GOES} soft X-ray fluxes in two channels, 1--8~\AA (red) and 0.5--4~\AA\ (blue), that include the contributions of both flares combined. Panel~(b) shows the spatially resolved AIA 94~\AA\ light curves of the two flares individually. The peak intensity of the secondary flare (blue) is about 10\% that of the main flare (red). If we assume that the AIA 94~\AA\ intensity is proportional to the {\it GOES} X-ray flux, this suggests that the secondary flare is on the order of {\it GOES} C5 class. Also note the 25~minute delay of the secondary flare from the onset of the primary flare. 

We selected four cuts for the space\,--\,time analysis, as shown in Figure~\ref{2flares:fig}(c): Cuts~0 and 1 are along the open funnel structure rooted at the main and secondary flare locations, respectively, which we use to capture the signatures of the outward propagating QFPs; Cut~2 traces the trans-equatorial loops connecting the two flares and is used for catching trapped modes, and Cut~3 is across the loops, with which we track both trapped modes and propagating QFPs. The resulting space--time plots are shown in Figures~\ref{spacetime:fig}(c)--(f). These are running ratio plots, i.e., each column of the space--time image is divided by the previous column of an earlier time frame in order to highlight subtle, rapid changes. This also helps minimize the effects of diminishing (on average) coronal emission with height that results primarily from gravitational stratification of the density. 

In general, evidence in emission variability due to QFPs is captured in all four cuts, especially in Cuts~0--2, and appears as recurrent, steep, narrow stripes. In the open-funnel Cuts~0 and 1 (see \figs{spacetime:fig}(c)--(d)), the stripes show mainly positive slopes, corresponding to outward propagating waves. The waves from the main flare (Cut~0) are particularly strong, traveling to more than $500 \arcsec$ (370~Mm) from the flare, and lasting for \emph{more than two hours}. This is a long duration compared to other features variability timescales, and in fact, the longest among all reported QFP events so far. In comparison, the open-funnel QFPs from the secondary flare travel only about a third of that range and only appear sporadically in time, but overall they are equally long lasting. 

More interesting are the waves captured in the closed loop Cut~2 (see \figw{spacetime:fig}(e)), where waves traveling in opposite directions are both present as evident from positive and negative slopes in the space\,--\,time diagram, and of comparable amplitudes in intensity variations (about 2\%--4\%). Although waves originating from each flare site appear upon the onset of its source flare, the counter-propagating waves are most evident during the period of 13:00-14:00~UT, as can be also seen in an enlarged view in Figure~\ref{model_vs_obs:fig}(b). These waves have a broad range of periodicities with typical periods falling in the 2--3~minutes range. These counter-propagating QFPs seem to be independently generated by their respective source flares, because of their tight temporal correlations with their flare onsets, but we cannot entirely rule out the possibility of reflections of the same QFPs from the opposing footpoints of the loops. Note that evidence for QFPs propagating in opposite directions \citep[][see their Figure~3]{Liu11} and the reflection of a single fast-mode wave pulse \citep{KI15}, both in closed loops associated with a single flare within a single AR of  different events, have previously been reported. However, the large amplitudes and spatial extents, and the long duration of the clearly apparent counter-propagating QFPs between two flares situated in two ARs presented here for the first time are truly remarkable.

As to the determination of the phase speeds of these waves, we applied parabolic fits to selected wave signatures in the space\,--\,time plots to provide examples for the speed ranges. We found initial speeds ranging from about 1000 to 4000~km~s$^{-1}$. The highest speeds are found in the closed loops at early times when the QFPs waves can be unambiguously identified close to the source flares. Comparably smaller initial speeds are seen later and at locations further away from the source flare, partly because the rapid growth of the flare arcade and its EUV intensity (shown as the shallow-sloped features near the top and/or bottom edges of the space--time plots in \figw{spacetime:fig}) prevent us from tracking individual QFP pulses close to the flare locations.  For each wave pulse we find that it generally decelerates as traveling away from the AR, which is consistent with the expectation of Alfv\'{e}n and fast-mode speed decreasing with height from the 3D MHD model of the bipolar AR (see, e.g., Figure~\ref{init_2dcut:fig}(b)). Details of the wave kinematics can be seen in the enlarged space\,--\,time plots in the left panels of \figw{model_vs_obs:fig}.

In addition to propagating QFPs, 
trapped modes are also evident, appearing as damped transverse kink oscillations in the space\,--\,time plot of \figw{spacetime:fig}(f) from the vertical Cut~3. To quantify the damped oscillations, we identified the average positions%
\footnote{Note that, unlike few simple cases \citep[e.g.,][]{Liu12,LO14}, in general, multiple loop strands participate in transverse oscillations, with various phase delays among them (see other examples, \citet{AS11}, their Figure~2), and and with each strand exhibiting temporally varying EUV intensities due to density and/or temperature, and line of sight integration effects. These effects add to observational ambiguities for tracking loop oscillations.}
of the oscillating loops from the sinusoidal features shown here and removed the gradual trend of the non-periodic displacement by subtracting a running smoothed version. The resulting positions were then fitted with a damped sine function of time, as shown in \figw{model_vs_obs:fig}(c), which indicates a period of $P= (21 \pm 2)$~min, a damping timescale $\tau= (6.2 \pm 1.6) P$, an initial amplitude of $A_0= (16 \pm 2)$~Mm, and an initial oscillation speed of $v_0= (79 \pm 9) \,\, {\rm km}\,{\rm s}^{-1} $. This result will be used for coronal seismological analysis to infer the magnetic field strength, as detailed in Section~\ref{CS:sec} below.



These kink oscillations are mostly likely triggered by the combination of the initial CME eruption and continuous QFPs launched from the flares. The damping of the oscillations could result from various processes, such as phase mixing, resonant absorption, and leakage of the waves \citep[e.g.,][]{OA02,MO08,SO10,Sel11a}, and we believe that leakage is the best candidate for the damping in this case, since we do not find strong evidence of heating expected to accompany wave dissipation processes.
%
%


Following \citet{Liu11}, we estimate the energy flux carried by the QFP waves
with the kinetic energy flux of the perturbed plasma in the WKB approximation,
 $ E = \rho (\delta v)^2 v_{\rm ph} /2
     \geq \rho ( \delta I / I)^2 v_{\rm ph}^3 / 8,  $ 
where we assume that the observed intensity variation $\delta I$ results from
density modulation $\delta \rho$ and 
use $\delta v/v_{\rm ph} \geq \delta \rho / \rho = \delta I / (2I)$ for magnetosonic waves since $I \propto \rho^2$ for EUV line emission intensity.  
This can be written in a normalized form
\begin{equation}
 E \geq  \left({n_e \over 10^8 \pcmc}\right)
         \left({ \delta I / I \over 1\% }\right)^2 
         \left({ v_{\rm ph} \over 1000 \kmps }\right)^3 
          [2.5\E{3} \ergs \pcms \ps]
\label{energy_eq}
\end{equation}
%
Here we take typical values measured at the apex of the loop Cut~2,
i.e., a phase speed of $v_{\rm ph} = 2000 \kmps $, a relative intensity perturbation of $\delta I / I = 3 \%$,
and a number density of $n_e \gtrsim 10^8 \pcmc$ estimated with differential emission measure (DEM) inversion for the large-scale corona (see Section~\ref{CS:sec}).
We then arrive at a lower limit of the energy flux of $E \gtrsim 1.8\E{5} \ergs \pcms \ps$.
Considering that the cross-section of the loops here has expanded substantially from the coronal base,
where the measured QFP speeds are also higher by about a factor of two, we estimate that
the corresponding energy flux there (i.e., near the source flare) would be at least one to two orders of magnitude higher.
Such energy fluxes are more than sufficient for heating the local AR corona \citep{WN77}, hence the importance of identifying and studying these waves in the solar corona.
\section{MHD Model}
\label{mhd:sec}
While the high cadence, high resolution AIA EUV images in general provide valuable information for the identification of QFPs, it is well known that such data unavoidably suffer from observational ambiguities, such as the line-of-sight projection effect, the unknown magnetic field strength, the true 3D geometry and the density structure. Use of observations alone may be insufficient to uniquely identify the MHD wave modes and their couplings. Such observational limitations thus necessitate supporting 3D MHD modeling, presented below, providing the means to disentangle various physical factors that contribute to the same observed feature and help identify the various MHD wave modes.  Moreover, by including several scenarios of wave sources and magnetic geometry in the model, we can infer the importance and effects of various parameters and make the numerical study more general with potential applications to other observations. Note that we do not expect exact one-to-one match between the observational and modeling results, because of the adoption of a simple bipolar magnetic geometry with idealized boundary conditions, but this approach, without loss of generality, does capture essential physics of MHD wave processes. Thus, the 3D MHD model provides realistic mode coupling and includes nonlinearity, lacking in more simplified linear analysis.



For this purpose, we solve the nonlinear, resistive 3D MHD equations with the standard notation for the variables given in a flux-conservative form
\begin{eqnarray}
&&\frac{\partial\rho}{\partial t}+\nabla\cdot(\rho\mbox{\bf V})=0,\label{cont:eq}\\
&&\frac{\partial(\rho\mbox{\bf V})}{\partial t}+\nabla\cdot\left[\rho\mbox{\bf V}\mbox{\bf V}+\left(E_up+\frac{\mbox{\bf B}\cdot \mbox{\bf B}}{2}\right)\mbox{\bf I}-\mbox{\bf BB}\right]=-\frac{1}{F_r}\rho\mbox{\bf F}_g,\label{mom:eq}\\
&&\frac{\partial\mbox{\bf B}}{\partial t}=\nabla\times(\mbox{\bf V}\times\mbox{\bf B})+\frac{1}{S}\nabla^2\mbox{\bf B},\label{ind:eq}\\
&&\frac{\partial(\rho E)}{\partial t}+\nabla\cdot\left[\mbox{\bf V}\left(\rho E+E_up+\frac{\mbox{\bf B}\cdot \mbox{\bf B}}{2}\right)-\mbox{\bf B}(\mbox{\bf B}\cdot\mbox{\bf V})+\frac{1}{S}\nabla\times\mbox{\bf B}\times\mbox{\bf B}\right]\nonumber\\	
&&=\frac{1}{F_r}\rho\mbox{\bf F}_g\cdot\mbox{\bf V}-n^2\Lambda(T)+\nabla_{||}\cdot(\kappa_{||}\nabla_{||} T)+H_{in}.\label{ener:eq} 
\end{eqnarray}
In the above equations the total energy density $\rho E=\frac{p}{(\gamma-1)}+\frac{\rho V^2}{2}+\frac{B^2}{2}$, the adiabatic index $\gamma$ 
 (an empirical polytropic index value of $\gamma=1.05$ is used in the present study),
the Alfv\'{e}n speed $V_A=B_0/\sqrt{4\pi\rho}$, the number density $n$, the optically thin radiative loss function $\Lambda(T)$ \citep{RTV78}, the empirical heating function $H_{in}$, the gradient parallel to the magnetic field $\nabla_{||}=\frac{\mbox{\bf B}}{\left|\mbox{B}\right|}\cdot\nabla$, and $\kappa_{||}$ the heat conduction coefficient parallel to $\mbox{\bf B}$ based on \citet{SH53}. 
The effect of the heat conductivity term is small in the present case, where the dominant wave mode is a fast wave, with shorts periods compared to the conductive dissipation time. Moreover, the present choice of an empirical polytropic index diminishes the effects of the source terms in the energy equation since the plasma is nearly isothermal with empirical value of $\gamma$ close to unity.

The normalizations in equations~(\ref{cont:eq})--(\ref{ener:eq}) are given by  $r\rightarrow r/L_0$, $t\rightarrow t/\tau_A$, $v\rightarrow
v/V_A$, $B\rightarrow B/B_0$, $\rho\rightarrow \rho/\rho_0$, and
$p\rightarrow p/p_0$, where $L_0=R_s/10$ and $R_s$ is the solar radius, $ \tau_A
= L_0/V_A$ is the \Alfven  time, $B_0$ is the background magnetic field, $\rho_0$ is the background density,
and $p_0$ is the pressure in the corona at the base of the active region. In the present study explicit viscosity is neglected. Other physical parameters are the Lundquist number $S=L_0V_A/\eta$, where $\eta$ is the resistivity  (in the present study we set  
$S=10^4$ and the resistivity has negligible effect on the waves), the Froude number  $F_r=V_A^2 L_0/(GM_s)$, where $G$ is the gravitational constant
and $M_s$ is the solar mass, and the Euler number $E_u=p_0/(\rho_0
V_A^2)=C_s^2/\gamma V_A^2$, where $C_s$ is the sound speed. 	
In  the present model we set $B_0=100$ G, $n_0=1.38\times 10^9$ cm$^{-3}$, $T_0=10^6$ K, which results in $V_A=5.87\times10^3$ km s$^{-1}$,  $\tau_A=11.9$ s, and $C_s=128$ km s$^{-1}$, and $F_r$=18.2. With these normalizations, the thermal to magnetic pressure ratio $\beta_0=8\pi p_0/B_0^2=2 E_u=9.6\times10^{-4}$. The equations are solved in a Cartesian geometry with the second order modified Lax-Wendroff method with a fourth order stabilization term \citep[e.g.,][]{Ham73} on a $256^3$ grid with a convergence test with double resolution grid. For improved stability the   solutions are obtained as nonlinear perturbations with respect to an initial state subject to the boundary conditions described  below. The numerical code used in the present study is an extension of the code NLRAT developed to study waves in coronal ARs and loops and tested in many previous studies \citep[e.g.,][]{OT02,TO04b,Ofm07,MO08,Ofm09,SO09,SO10,Sel11a,Sel11b,Ofm11,Ofm12,Wan13,Ofm15,POW18}. 

\subsection{Initial State and Boundary Conditions}
\label{init:sec}

\begin{figure}[ht]
\center
\includegraphics[width=10cm]{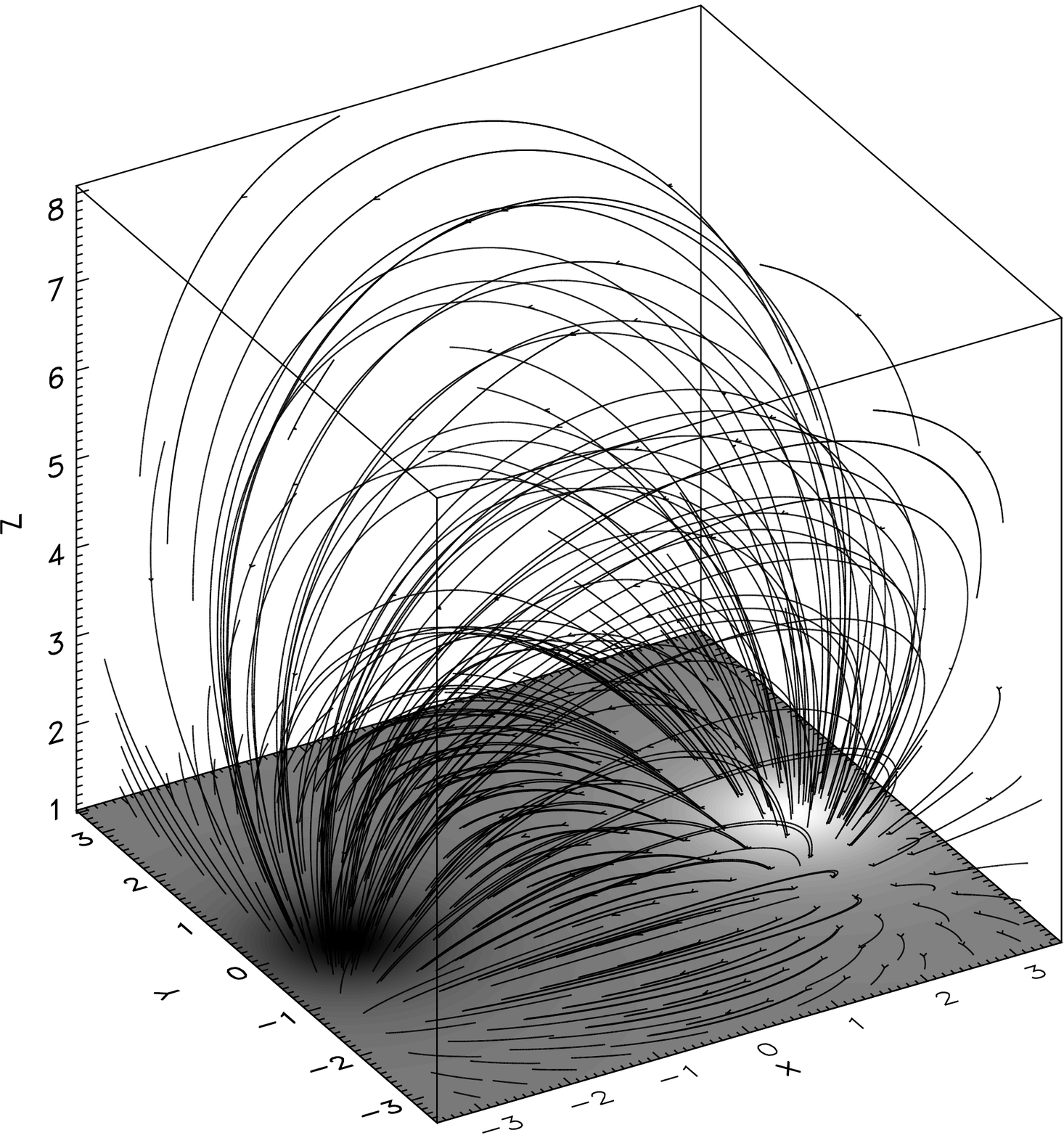}
\vspace{-1cm}\caption{The initial, bipolar model AR magnetic field configuration for a dipole separation of $d=2.5$. An animation of the 3D magnetic field due to the impact of the QFPs is available online.}
\label{B3d:fig}
\end{figure}

The AR is modeled in a Cartesian geometry, initialized with a bipolar magnetic field produced by a dipole located below the `surface' \citep[e.g.,][]{COS89,OT02}. The magnitude of the normalizing magnetic field is constrained to produce a 2500 km~s$^{-1}$ fast magnetosonic speed at the lower boundary of the propagating QFPs, which is in the range of observed speeds for our observationally guided model. The parameters of the dipole, such as the separation of the magnetic poles and the location of the dipole source, determine the details of the magnetic geometry of the model AR \citep[e.g.,][]{Ofm15}. In particular, we set the aspect ratio between the height and the footpoint separation of the magnetic loops to 0.7 to approximately match the oscillating loops in the dual M5.0 flare event observed with AIA in 171~\AA\ on 2013 May 22.  This is achieved by using the dipole separation $d=2.5$ in Equations~(5)-(10) of \citet{OT02} (see Figure~\ref{B3d:fig}). We vary the values of $d$ and other model parameters in several runs described in Section~\ref{num:sec} below. In the model we choose an identical periodicity of 3~minutes, and vary the separation of the QFP wave sources,  the magnetic topology, and amplitude in a limited parametric study that elucidates the main effects of these parameters. More detailed parametric studies are subjects for future work.

\begin{figure*}[ht]
\center
\includegraphics[width=15cm]{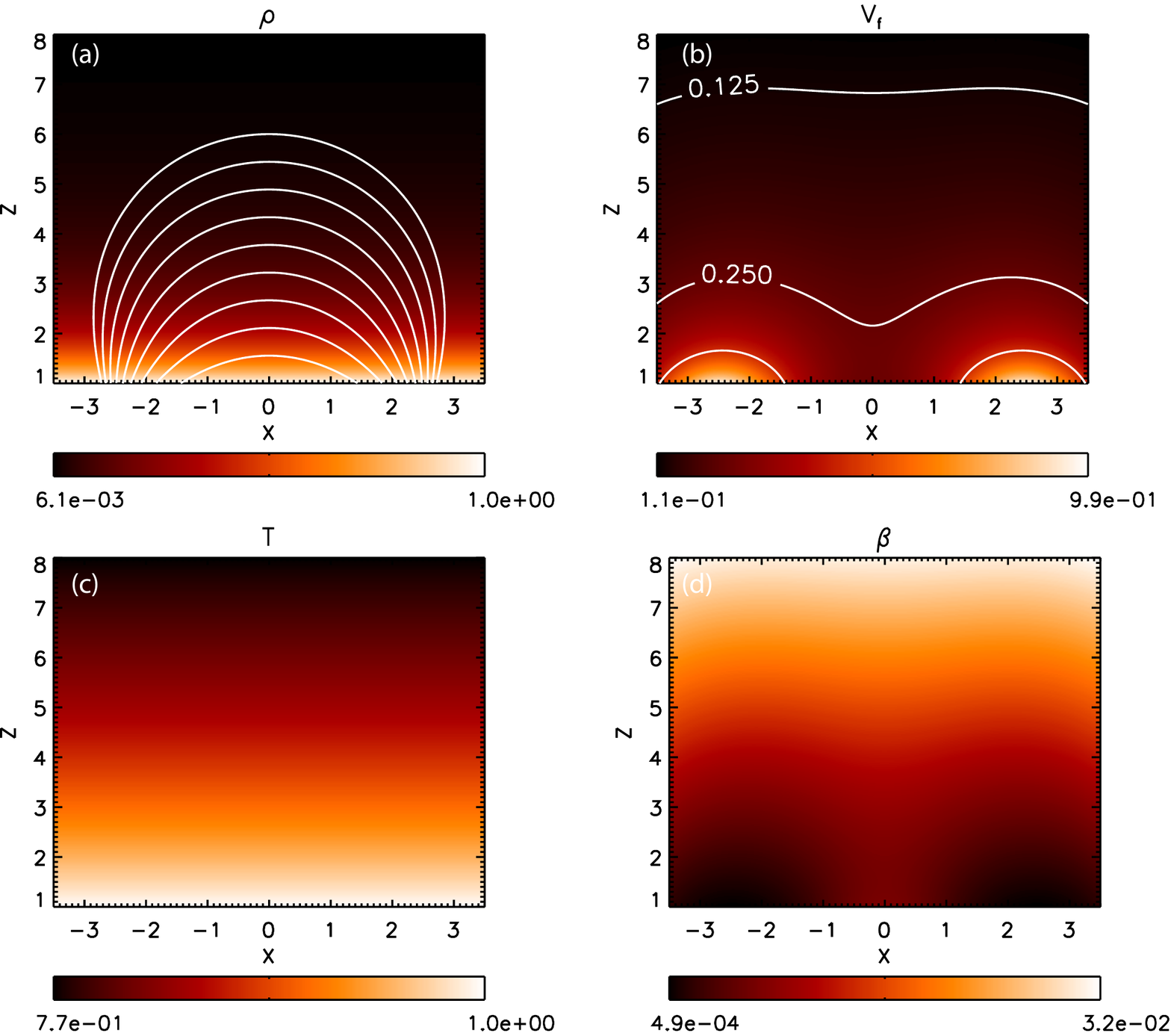}
\vspace{-3cm}\caption{The initial state of the model AR variables, showing a cut in the $x-z$ plane at $y=0$ for the 3D dipole magnetic field with $d=2.5$. (a) The initial density, $\rho$, and selected magnetic field lines. (b) The initial fast magnetosonic speed, $V_f$. The color map and the contours show the spatial dependence of $V_f$. (c) The initial normalized background temperature, $T$. (d) The initial values of plasma $\beta$.}
\label{init_2dcut:fig}
\end{figure*}

The initial normalized equilibrium gravitationally stratified polytropic density is introduced into the potential field, given by
\begin{eqnarray}
\rho_0=\left[1+ \frac{\gamma-1}{\gamma}\frac{H}{10}\left(\frac{1}{1+0.1(z-1)}-1\right)\right]^\frac{1}{\gamma-1},
\label{rho:eq}\end{eqnarray} where $H=2k_BT_0L_0/(GM_sm_p)$ is the normalized gravitational scale height, and $k_B$ is the Boltzmann constant. This expression is an approximate model of spherical, gravitationally stratified density equilibrium with length scale $L_0=0.1R_\odot$. The initial normalized temperature is related to the normalized density through the polytropic relation $T=T_0\rho^{\gamma-1}$. With the above magnetic and density structure the initial fast magnetosonic speed \begin{eqnarray}
&&V_f(x,y,z)=\left\{\frac1{2}\left[V_A(x,y,z)^2+C_s^2+((V_A(x,y,z)^2+C_s^2)^2-4C_s^2V_A(x,y,z)^2\cos^2\theta\right]^{1/2}\right\}^{1/2}
\end{eqnarray} is nonuniform, where $\theta$ is the angle between the magnetic field and the wave vector. This initial state is an equilibrium polytropic atmosphere without the effects of explicit heating or cooling (see, Figure~\ref{init_2dcut:fig}).
\vspace{-3cm}

We use the following time-dependent boundary conditions at the lower coronal boundary (defined at $z=1$) to launch the magnetosnoic waves at two locations (referred to as subscripts 1 and 2),
\begin{eqnarray}
&&v_x(x,y,t)=\sum_{i=1,2} v_{x,i}\,{\rm sin}(\omega_i t+\phi_i)e^{\frac{-(x-x_i)^2-(y-y_i)^2}{w_i^2}}
\end{eqnarray} where $v_{x,1,2}$ are the amplitudes of the driving sources, $w_{1,2}$ are the half-widths of the source regions at the lower boundary, $\omega_{1,2}$ are the frequencies, and $\phi_{1,2}$ are the phases. In all cases we use $y_i=0$, and values of $d$, $\omega$, $x_i$, and $v_{x,i}=V_0$ are given in Table~1. The choice of these wave parameters is motivated by the observations described in \sect{obs:sec} with possible relevant variations. Since the fast magnetosonic speed, $V_f$, is nonuniform in the bipolar AR, the location of the sources at the base determines the direction of propagation of the waves due to the refraction of the wavefronts governed by the $V_f$ gradients. Thus, placing the sources of the waves near the opposite poles results in counter propagating wave trains. In general, the model parameters with subscript 1 and 2 are independent, and we have studied several scenarios with different locations of the two sources guided by the observed event. The remaining boundary conditions at $z=1$ are $v_y=v_z=0$, with $\mbox{\bf B}$ components, density $\rho$, and pressure are extrapolated from the interior points. The boundary conditions on the top and sides of the computational domain are open, allowing the escape of the waves with negligible reflection. 
%
\begin{table}
\caption{Summary of key model parameters for the five cases of 3D MHD simulation.}	
\hspace{1cm}\begin{tabular}{cccccc}
\hline
	Case \#  & $V_0$ [$V_A$] & $\omega$  &  $d$ & $x_1$ & $x_2$  \\ \hline
             1       &  0.02                    &      0.4189  & 2.5 & $-2.0$  & 2.0  \\
            2       &  0.02                     &      0.4189  & 2.5 & $-2.4$  & 2.4  \\
            3        & 0.02                     &    0.4189  & 2.5 & $-2.0$ & 2.4  \\
          4         & 0.02                     & 0.4189      &   2.0  &   $-2.0$      &   2.0  \\
          5        & 0.04                     & 0.4189      & 2.5   &  $-2.4$ & 2.0 \\
\hline
\end{tabular}

\label{cases:tab}
\end{table}

\section{Numerical Results}
\label{num:sec}
\begin{figure*}[ht]
\center
\includegraphics[width=13cm]{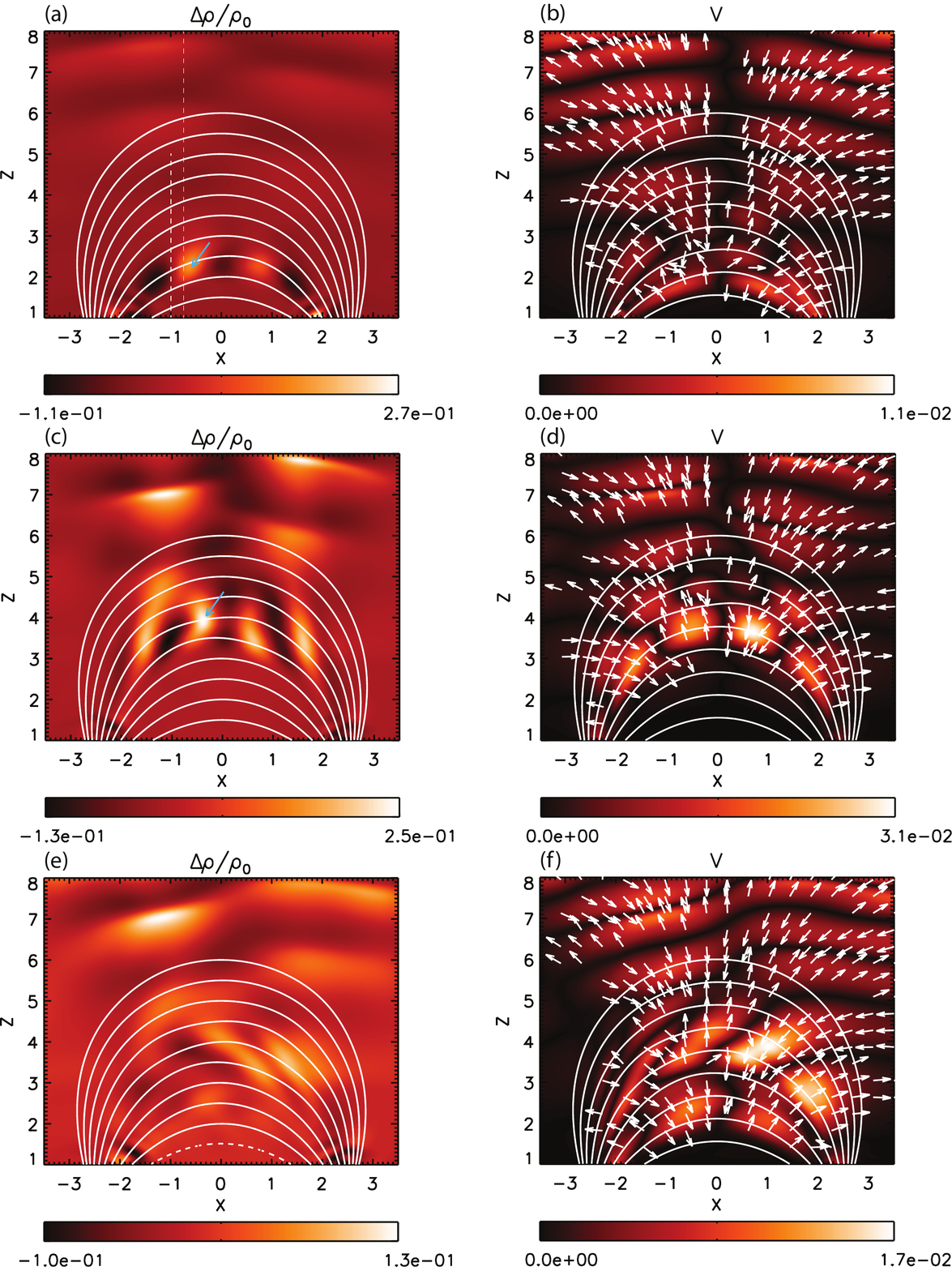}
\caption{Counter-QFP produced density ($\Delta \rho/\rho_0$, left) and velocity ($v$, right) perturbations in the $x$\,--\,$z$ plane at $y=0$, overlaid with magnetic field lines and velocity direction arrows (right panels only). From top to bottom are three simulation cases: (a) and (b) Case~1 at $t=119\tau_A$, (c) and (d) Case~2 at $t=98.4\tau_A$, (e) and (f) Case~3 at $t=98.3\tau_A$. The dashed line styles mark the locations of the space-time cuts of the model results in Figure~\ref{model_vs_obs:fig}(d)-(f) below. The cyan arrows mark some of the compressions associated with the trapped slow mode wave. An animation of panel (e) is available online.}
\label{vb_2qfp:fig}
\end{figure*}

\begin{figure*}[ht]
\center
\includegraphics[angle=270,width=15cm]{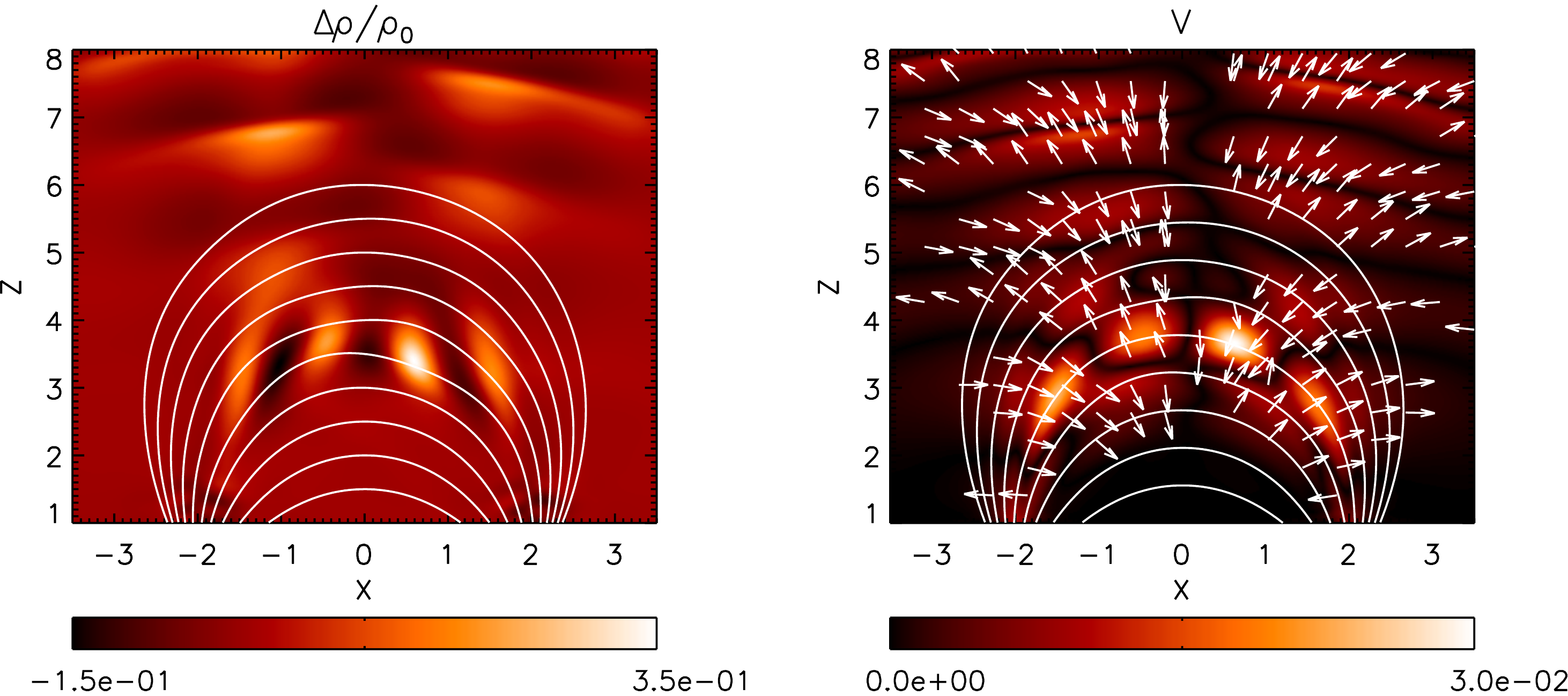}
\vspace{-1cm}\caption{Same as Figures~\ref{vb_2qfp:fig}(a) and (b) (Case~1, $d=2.5$), but for Case~4 ($d=2$) at $t=98.5\tau_A$.}
\label{vb_2qfp_d2:fig}
\end{figure*}

In Figures~\ref{vb_2qfp:fig}\,--\,\ref{vb_2qfp_d2:fig} we present the numerical results of several cases of QFPs excited at two locations with the parameters summarized in Table~1. In Figure~\ref{vb_2qfp:fig} the snapshots of the relative density perturbations, defined as $\Delta \rho/\rho_0$ (where $\Delta \rho\equiv\rho-\rho_0$), the velocity and the magnetic field lines are shown for Cases~1\,--\,3 in the $x-z$ plane at $y=0$ of the model AR. The times are $t=119\tau_A$ (Case~1), $t=98.4\tau_A$ (Case 2), and   $t=98.3\tau_A$ (Case~3). At these times the oscillations are in or nearly at the steady state. 	
From the animations and the space-time analysis (see below) it is evident that the QFPs excited at opposite footpoints of magnetic loop produce propagating as well as trapped waves. The propagating waves crossing the AR bipolar magnetic field structure are evident in the density and velocity perturbations, as expected for fast magnetosonic waves. These waves leave the domain at the top and side boundaries. However, the two QFPs also excite trapped modes, as evident in the velocity and density perturbations inside the loops of the AR magnetic domain. The density structure is modified by the nonlinear effect of magnetosonic wave pressure in these magnetic loops, consistent with previous observations and modeling studies in a more simplified cylindrical  coronal loop geometry \citep[e.g.,][]{TO04b} and in bipolar magnetic  geometry \citep{MO08,SO09,SO10,Sel11a,Sel11b,Ofm12}. The trapping effect is strongest for the cases where waves are excited at exactly conjugate footpoints of the same magnetic loop, where the height of the apex of the excited loops is directly related to the footprint separation. When the locations of the QFP wave excitation is non-symmetric with respect to the AR magnetic structure (Case~3), the location of the trapped waves is shifted towards the location of the more distant wave source (see, Figures~\ref{vb_2qfp:fig}e-\ref{vb_2qfp:fig}f). Nevertheless, both, propagating and trapped waves are evident in Case~3 as well. 

In order to examine the effects of the dipole field topology, such as the aspect ratio between the pole separation and the height of magnetic loops, in Case~4 we use the same parameters as in Case~1, but with smaller separation of the magnetic poles, $d=2$ (see Figure~\ref{vb_2qfp_d2:fig}). The QFP waves were excited at the same locations, i.e., $x_{1,2}=\pm 2.0$ with identical periods and amplitudes. We find that the main difference between Cases~1 and 4, is the increased height of the magnetic loops that carry the trapped mode in Case~4 compared to Case~1, due to the difference in topology of the dipole magnetic field with  $d=2.0$ compared to $d=2.5$. Specifically, the wave sources in Case~4 are located at the footpoints of comparably taller magnetic loops than those in Case~1. This changes the relative spatial distribution of the fast-mode speed $V_f$, whose spatial gradient governs the propagation direction of fast-mode magnetonoic waves, and these waves propagate obliquely to the direction of the magnetic field lines. In addition there is a  change in the wavelength and relative amplitude of the propagating components due to the change of the fast magnetosonic speed dependence with hight, with more rapid decrease of $B$ with height, compared to the decrease of the density with height. 

In the Appendix, we discuss the temporal evolution of various physical variables for the QFP-wave driven perturbations at a fixed location in the 3D model AR. Specifically, Figures~\ref{bvnTt1:fig}--\ref{bvnTt5:fig} show the details of the phase relations between these variables, which help the identification of the dominant wave modes at this location. 

\subsection{Comparison to observations}

\begin{figure*}[ht]
\center
\includegraphics[angle=90,width=18cm]{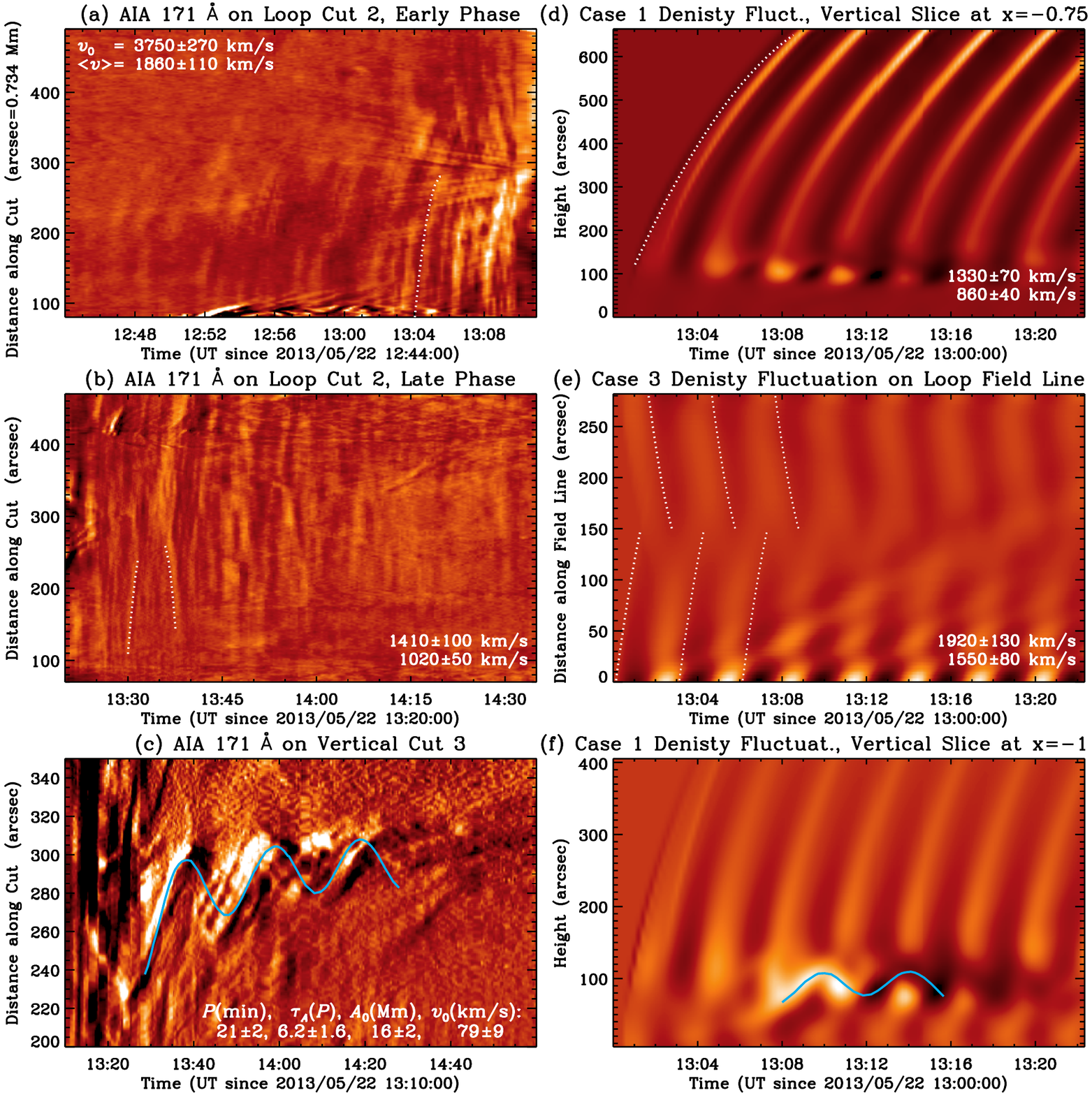}
\caption{
Comparison of space--time plots from the observations (left) and simulations (right), showing  
the deceleration of QFPs with distance (top panels), 
counter-propagating waves (middle panels),
and trapped kink oscillations (bottom panels). See text for details.
}\label{model_vs_obs:fig}
\end{figure*}
%
%

 In Figure~\ref{model_vs_obs:fig} we compare space\,--\,time plots of the observed EUV intensity from the 2013 May 22 QFP event described in Section~\ref{obs:sec} and of the plasma density perturbations from the 3D MHD simulations of the counter-propagating QFPs in the model AR, since the density is directly related to the emission measure and thus the EUV intensity. To make direct comparison with observations, we applied the same parabolic fits to simulated wave trajectories. {bf We show the results of Case~1 that emphasizes the symmetric interactions of the counter~QFPs, and Case~3 to show the effects of slightly non-symmetric QFPs excitation.} In the left panels of Figure~\ref{model_vs_obs:fig} we show the AIA 171~\AA\ intensity running ratio plots from the loop Cut~2 connecting the two neighboring flares during (a) the early phase and (b) the late phase, and (c) from the vertical Cut~3 
(i.e., enlarged subregions of Figures~\ref{spacetime:fig}(e) and (f)). In the right panels of Figure~\ref{model_vs_obs:fig} we show 
 the density perturbations space--time plots from simulations:
(d) from a vertical cut located at $(x, \,y)=(-0.75,\, 0)$ in Case~1,
(e) along a magnetic field line in the $x$--$z$ plane passing through $(x, \,z)=(0, \,1.5)$ in Case~3,
and (f) from a vertical cut located at $(x, \,y)=(-1,\, 0)$ in Case~1. The cuts in the model are indicated with dashed line styles in Figure~\ref{vb_2qfp:fig}. Panels (d) and (e) are running ratios, while (f) is a running difference to highlight the kink oscillations at low heights. The low-laying loop was chosen to show the counter-QFPs since it is below the trapped mode wave region. 
The dimensionless distance has been converted to physical distance in units of arcseconds. The white dotted lines in the top and middle panels are parabolic fits to selected wave trajectories, with the initial and average speeds indicated for the first fit of each panel showing similar magnitudes. The cyan curve in (c) is a fit to the transverse oscillations of the trans-equatorial loops 
with a damped sine function, with the best-fit parameters indicated at the bottom of the panel.
The cyan curve in (f) indicates transverse oscillations in the simulation. 
 
We found remarkable agreement between the observed and simulated space--time plots.
In particular, the dominant propagating waves from the main flare during the early phase shown in Figure~\ref{model_vs_obs:fig}(a)
resemble those simulated upward propagating waves shown in Figure~\ref{model_vs_obs:fig}(d), both in the form of steep stripes with the QFP periodicity and fast-mode wave speed, as well as considerable deceleration.
The counter-propagating waves between the two flares during the late phase shown in Figure~\ref{model_vs_obs:fig}(b) are also quite similar to those
captured on a closed magnetic field line of the dipole in Figure~\ref{model_vs_obs:fig}(e), where the free parameters of the 3D MHD model are the magnitudes of the magnetic field, density, and temperature at the coronal base (see Section~\ref{mhd:sec}). Note that the shallow-sloped, recurrent features in Figure~\ref{model_vs_obs:fig}(e) with typical speeds of $\sim$$140 \kmps$ are due to slow mode waves launched from the QFP source region.

The model results are in excellent agreement with the observations in terms of wave propagation morphology phase speed range and couplings, and also exhibit transverse oscillations of qualitatively similar periodicities and intensity variations. 
From our analysis of the modeling results, it is evident that the oscillations are due to the trapped kink mode in the loop structure, while the intensity variation is a result of a slow magnetosonic wave driven nonlinearly by the QFPs. 

In summary, both the trapped oscillations and propagating QFPs are evident in the observations and simulations, allowing the confirmation of their physical nature due to the various MHD modes, with additional evidence of the slow mode in the intensity variation of the observed oscillating loops.

\subsection{Coronal Seismology}
\label{CS:sec}

In order to infer the typical magnetic field strength, we apply coronal seismology \citep[e.g.][]{NV05} by using the observed properties of the oscillating loops (including their period and density) in the 2013 May 22 event.

We identify the most evident oscillating loop (as marked by the arrow in Animation~2) that corresponds to the transverse oscillations shown in Figure~\ref{model_vs_obs:fig}(c). This loop has a width of $w=(1.14 \pm 0.11) \times 10^9 \cm $ and a length of $L=(5.41 \pm 0.54) \times10^{10} \cm$, inferred by fitting the AIA~171~\AA\ loop in the plane of the sky with a curve fit and assuming a $10^\circ$ longitudinal span along the line-of-sight (as indicated by the EUV images). 

We estimate the plasma density of this loop using the DEM inversion technique of \citet{Che15}. To do so, we constructed DEM maps of this region in multiple temperature bins in the $\log T=5.5$ to 7.5 range. We then selected five loop segments of $20 \arcsec$ long near the loop apex spaced by $20 \arcsec$ to $50 \arcsec$. We obtained the spatially averaged DEM of each segment, DEM$_0$, and of its immediate neighborhood above it of a similar size, DEM$_{\rm bkg}$, which was used as a background to be subtracted to obtain the DEM for the loop itself, ${\rm DEM}_{\rm loop} = {\rm DEM}_0 - {\rm DEM}_{\rm bkg}$. The loop density is thus given as $n = \sqrt{ [\int {\rm DEM}_{\rm loop} d T] / w}$. We use the average of the densities of five segments as the final estimate for the loop density and their standard deviation plus a 10\% systematic error as the uncertainty, $n_i=(6.34 \pm 0.80) \times10^8 \pcmc$. Likewise, the background density is estimated from $n = \sqrt{ [\int {\rm DEM}_{\rm bkg} d T] / R_s}$ at $n_o= (9.01 \pm 1.24) \times10^7 \pcmc$, by assuming a line-of-sight integration length of $R_s$. This gives a loop density contrast of $n_o/n_i=0.14\pm0.04$.


Using the oscillation period ($P=21\pm 2 {\rm min} = 1260\pm 120 \s$) from Figure~\ref{model_vs_obs:fig} and the values of the density and loop length inferred above, we obtain the kink speed $C_k$ as
\begin{eqnarray}
C_k=2L/P=2\times 5.41\times10^{10}/1260 \ \cm \ps =8.59\times10^7\ \cm \ps =859\ \kmps, 
\end{eqnarray}
using the fundamental mode of oscillation. Since the kink speed depends on the Alfv\'{e}n speed and on the loop density contrast as \citep[e.g.,][]{NV05}
\begin{eqnarray}
C_k=V_A\sqrt{\frac{2}{1+n_o/n_i}}, 
\end{eqnarray} we derive the value of the Alfv\'{e}n speed inside the loop $V_A=B/(2\pi\rho_i)^{1/2}=651 \kmps$ and using the loop density $n_i$ obtained above we find the magnitude of the magnetic field $B=5.3$~G. The fast magnetosonic speed inside the loop can be obtained from $V_{f,i}=(V_A^2+C_s^2)^{1/2}=670 \kmps$ for a perpendicular propagation angle. Outside the loop the fast magnetosonic speed becomes $V_{f,o}=\left[V_A^2 (n_i/n_o)+C_s^2\right]^{1/2}=1744 \kmps$, consistent with the range of QFP propagation speeds that we find in the observations and in the 3D MHD model, as indicated in Figure~\ref{model_vs_obs:fig}. In the above analysis the uncertainties are on the order of 10\%.

The magnetic field intensity of the dipole in the 3D MHD model decreases rapidly with height in agreement with observed active region field structure, and with the present model normalization, we find $ֿB\approx 5$ G near the center of the 3D model AR, in agreement with the above coronal seismology result. However, the average magnetic field strength along the magnetic loop with its apex at the center of the model AR is substantially higher due to the contribution of the stronger magnetic field near the footpoints \citep[see also,][]{Ofm12,Ofm15}. Moreover, the high density contrast of the individual loop is not included in the simplified model. This is acceptable, because the main properties of 3D MHD modeling are already in agreement with observations, and the present study is not aimed at quantitative coronal seismology with the 3D MHD model.

\section{Conclusions and Discussion}
\label{disc:sec}
We report and analyze the first detection of counter-propagating QFP waves associated with two neighboring flares observed by \sdo/AIA in EUV on 2013 May 22. We find that the two flaring regions are connected with large-scale trans-equatorial coronal loops, and the waves propagate across the magnetic structures in opposite directions at large speeds of the order $>$1000 km s$^{-1}$. These observations provide evidence of counter-propagating fast magnetosonic wave trains, as well as their interaction and generation of trapped waves that are identified as damped kink modes in adjacent coronal loops. Using space\,--\,time analysis, we measure the properties of the various wave modes, such as the phase speeds in various directions, and study properties of the loop oscillations. We also apply coronal seismology using loop oscillations during the event, and determine the magnitude of the magnetic field in these loops with the density obtained from DEM analysis. The corresponding phase speed of the QFPs from coronal seismology analysis is in agreement with the observationally determined range of values. 


Using the observations as a guideline, we utilize our 3D MHD code to setup and model the bipolar AR and study the counter-propagating QFPs from the two flaring sites, simulated by the time-depended periodic boundary conditions.
We investigate the excitation, propagation, and interaction of the counter-propagating waves for various combinations of different excitation locations and bipolar magnetic geometries in a limited parametric study. 
We confirm the identification of fast-mode waves as the dominant modes for the propagating component, as well as damped kink mode and nonlinearly generated slow mode waves for the trapped component in the magnetic loops. We find that the 3D MHD model reproduces the main observed features of the counter-QFPs in the dual flare event on 2013 May 22 seen by \sdo/AIA  providing insights on linear and nonlinear MHD mode excitation, propagation and couplings. These results provide new evidence of the correlation of flares, QFPs, and the related trapped (kink mode) waves in AR coronal loops. 
	

In particular, when the excitation locations of counter-QFPs are symmetric with respective to the dipole, they lead to formation of both propagating fast-mode waves and standing kink mode waves due to trapping of the waves in closed loops. In addition, the magnetic wave pressure variation produces a slow magnetosonic wave, that eventually leads to the formation of a standing slow mode wave in the loop. However, the slow magnetosonic wave is energetically insignificant due to its small amplitude and low phase speed and the correspondingly small energy flux that is proportional to the sound speed. With similar amplitudes of the velocity perturbations, the QFPs carry most of the energy flux supplied by the flares due to their large fast magnetosonic phase speed in the low plasma $\beta$ corona. While the nonlinear effects due to the magnetic wave pressure are significant in producing some steepening of the wave fronts, and compressional waves, they are weakly nonlinear in all cases for the parameters relevant to the observed event.

We find that the asymmetric excitation of the counter-QFPs in the model is in best agreement with the observed event, where there is a 25~minute delay between the onset times of the two flares. Since Alfv\'{e}n and fast mode waves are coupled in inhomogeneous plasma the counter-QFPs could facilitate the onset of turbulent cascade and MHD wave dissipation in the lower corona. It is therefore important to note that our observations provide the first direct evidence of counter-propagating fast magnetosonic waves that can carry sufficient energy flux for coronal heating in the reported observation, similar to the flux needed in Alfv\'{e}n waves in low-$\beta$ coronal plasma, thus, supporting the scenario of counter-propagating waves leading to turbulence dissipation. While the QFPs can couple to the Alfv\'{e}n mode, the significant compressibility and the oblique propagation of QFPs with weak trapping is very different from the properties of pure Alfv\'{e}n waves in incompressible plasma, and some of the QFP energy flux can dissipate through other processes, such as compressive damping. 

 We note that thus far, since the beginning of the SDO mission in 2010,
only about a dozen QFP events have been reported with detailed analysis and about 100 more events have been identified in a statistical study \citep{LO14,Liu16}.
Almost all these events were associated with flares of various magnitudes.
If this relatively low detection rate represents their true occurrence rate on the Sun, 
this would imply that QFPs, as seen by SDO/AIA in EUV, are possibly not the primary
contributor to heating the global solar corona. On the other hand, we speculate
that if the hypothesis of nanoflares being responsible for coronal heating holds 
\citep[e.g.,][]{Par88}, and if most nanoflares produce QFPs that carry a significant fraction of their energy flux, 
then one would expect that dissipation of numerous small-scale QFPs can make significant contribution
to coronal heating. However, same as for nanoflares, such QFPs are below the direct detection threshold of 
the current instruments. Thus, the role of QFPs in coronal heating remains to be verified with future 
instrumentation of improved capabilities.

\acknowledgments The authors acknowledge support from NASA HGI grant NNX16AF78G. 
W.L. is also supported by NASA LWS grant NNX14AJ49G to PSI and NASA {\it SDO}/AIA contract NNG04EA00C to LMSAL. W.L. thanks Ineke De Moortel for helpful discussions. Resources supporting this work were provided by the NASA High-End Computing (HEC) Program through the NASA Advanced Supercomputing (NAS) Division at Ames Research Center. We thank Dr.~Xudong Sun for help with analysis of \sdo/HMI magnetograms that provided observational
support for the bipolar magnetic geometry adopted in our model. 
AIA is an instrument onboard the {\it Solar Dynamics Observatory}, a mission for NASA's Living With a Star program.
 
\section*{Appendix}

\begin{figure*}[ht]
\center
\includegraphics[width=15cm]{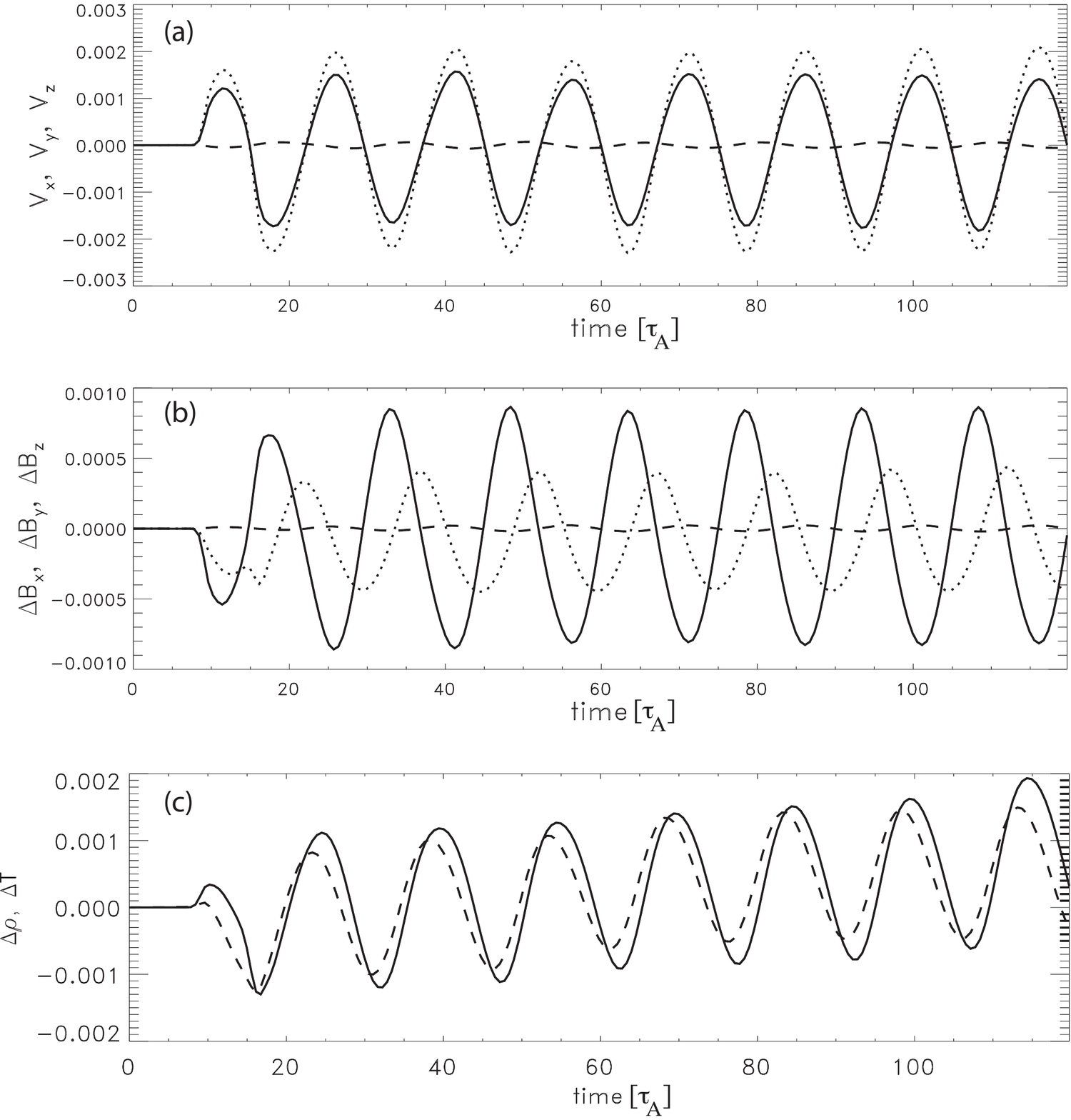}
\vspace{-2cm}\caption{The temporal evolution of physical variables at $x=1.2$, $y=0$, $z=3.5$ with symmetric excitation of QFPs at $x_{1,2}=\pm2.0$ (Case~1). The time is in units of $\tau_A=11.9$ s. (a) The velocity components $V_x$ (solid), $V_y$ (dashes), $V_z$ (dots). (b) The perturbed magnetic field components $\Delta B_x$ (solid), $\Delta B_y$ (dashes), $\Delta B_z$ (dots). (c) Density (solid) and temperature (dashes) perturbations, $\Delta \rho$ and $\Delta T$.}
\label{bvnTt1:fig}
\end{figure*}

\begin{figure*}[h]
\center
\includegraphics[width=15cm]{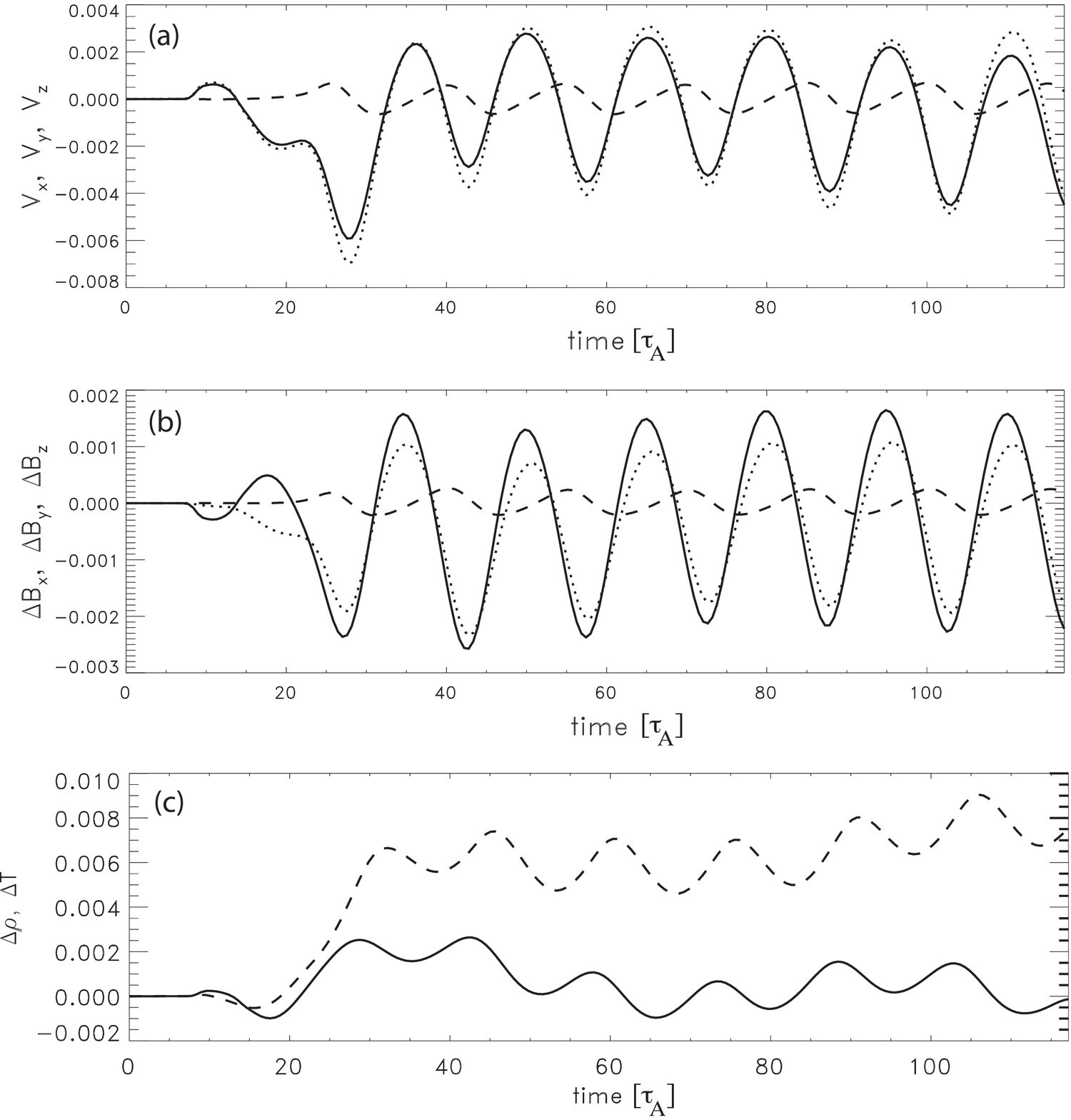}
\vspace{-2cm}\caption{Same as \figw{bvnTt1:fig}, but for Case~3 with non-symmetric excitation of QFPs at $x_1=-2.0$, $x_2=1.4$. (a) The velocity components $V_x$ (solid), $V_y$ (dashes), $V_z$ (dots). (b) The perturbed magnetic field components $\Delta B_x$ (solid), $\Delta B_y$ (dashes), $\Delta B_z$ (dots). (c) Density (solid) and temperature (dashes) perturbations.}	
\label{bvnTt2:fig}
\end{figure*}

\begin{figure*}[h]
\center
\includegraphics[width=15cm]{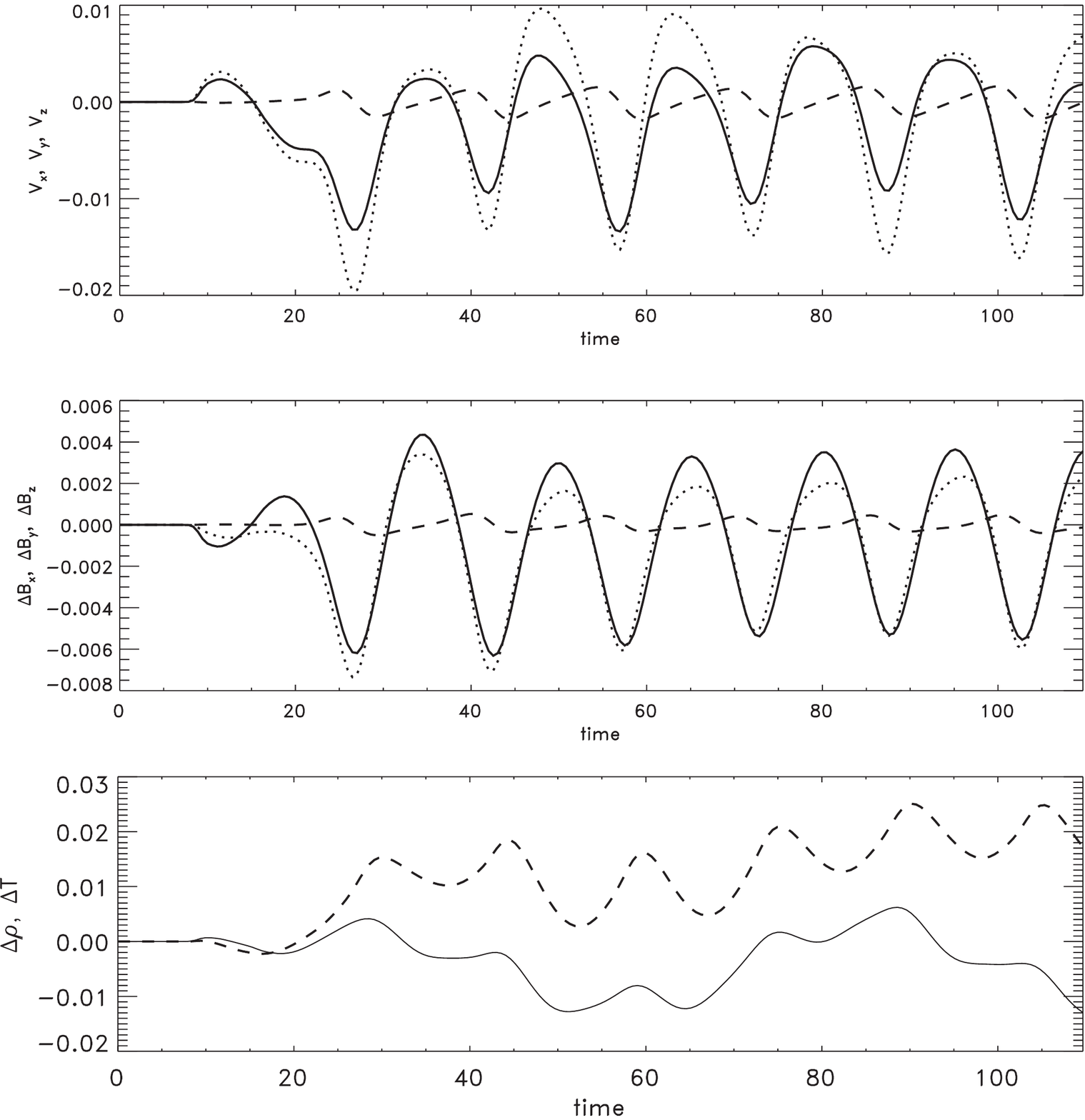}
\vspace{-2cm}\caption{Same as Figure~\ref{bvnTt2:fig} but for Case~5 with double the wave amplitude. (a) The velocity components $V_x$ (solid), $V_y$ (dashes), $V_z$ (dots). (b) The perturbed magnetic field components $\Delta B_x$ (solid), $\Delta B_y$ (dashes), $\Delta B_z$ (dots). (c) Density (solid) and temperature (dashes) perturbations.}
\label{bvnTt5:fig}
\end{figure*}

Here we show the temporal evolution of the QFP wave driven perturbations of various physical variables at a fixed point in the 3D model AR. The details of the phase relations between the variables, helping the identification of the dominant wave modes are evident in the temporal evolutions shown in Figures~\ref{bvnTt1:fig}--\ref{bvnTt5:fig}. The evolution at $x=1.2$, $y=0$, $z=3.5$ for Case~1, with QFPs excited at $x_{1,2}=\pm2.0$ is shown in Figure~\ref{bvnTt1:fig}.  The initial interval of $\sim10\tau_A$ without wave signal is the time for the perturbations to reach the `observational point'  in the model AR. The velocity components $V_x$ (solid), and $V_z$ (dots) oscillate in phase, while the $V_y$ (dashes) shows a quarter wavelength phase shift. While $V_x$ and $V_z$ belong to the fast magnetosonic wave component of the QFPs, the $V_y$ component is Alfv\'{e}nic, and is produced by the couplings in the non-uniform dipole field. Note that the perturbed magnetic field (with respect to the initial dipole) component $\Delta B_x$ is in anti-phase with $V_x$, and there is a quarter wavelength shift between $\Delta B_z$ and $V_z$. These phases are likely due to the interaction between the two counter-QFPs, since a quarter wavelength shift is typical for a standing wave mode, comparing to Figure~5 in \citet{Ofm11} for a single-source QFP wave.

The temporal evolution of the perturbed density $\Delta \rho$ and temperature $\Delta T$ (where the perturbations are with respect to the initial state) at $x=1.2$, $y=0$, $z=3.5$ for Case~1 are shown in in Figure~\ref{bvnTt1:fig}(c). It is evident that the oscillations of $\Delta \rho$ and $\Delta T$ are nearly in phase, where the small phase shift is due to the small effect of thermal conduction, similar to the phase shift in slow magnetosonic waves \citep[see, e.g.,][]{Owe09,Van11,Wan15}. In addition to the oscillations, there is a small gradual increase in the average background density and temperature due to the effects of the wave pressure on the background. 

In Figure~\ref{bvnTt2:fig} the temporal evolution of the variables at the same fixed point $(x,y,z)=(1.2, 0, 3.5)$ as in Figure~\ref{bvnTt1:fig} but for Case~3 are shown. The main effects of the non-symmetric double QFP excitations is the change in phase relations between the $z$ components of the velocity and the magnetic field, and the deviation from pure sinusoidal oscillations. Also, the initial transient state is evident up to $t\approx30\tau_A$ in velocity and magnetic field perturbations, and to somewhat longer time in density and temperature perturbations. Following this initial transient, the variables reach nearly steady oscillations.	
While the nonlinear effects are present in all cases, they are of small magnitude due to the low amplitude of the QFP velocity sources compared to the local fast magnetosonic speed (see~Table~1). In order to demonstrate the effect of the amplitude on the nonlinearity, in Case~5 we use $V_0=0.04$, doubling the amplitude of Case~3. Figure~\ref{bvnTt5:fig} shows the evolution of Case~5, where the stronger nonlinearity is most evident in the non-sinusoidal velocity fluctuations due to the waves, as well as in the magnetic field, density, and temperature perturbations. The effect of stronger wave pressure  results in larger perturbations in the density and temperature in this case, as evident by comparing the scales of the perturbations in Figures~\ref{bvnTt2:fig}(c) and \ref{bvnTt5:fig}(c).

\end{document}